\documentclass{ifacconf}

\usepackage{graphicx}      
\usepackage{natbib}        
\usepackage{color}
\usepackage{amsmath,mathrsfs}
\usepackage{amsfonts}
\newtheorem{remark}{Remark}
\usepackage{epstopdf}    
\usepackage{algorithm}
\usepackage{balance} 
\usepackage{multirow}
\usepackage{algorithmicx}
\usepackage{algpseudocode}
\usepackage{xfrac}
\usepackage{bm}
\usepackage{enumitem}
\usepackage{pgfplots}
\usepgflibrary{shapes}
\usepackage{tikz}
\usepackage{verbatim}
\usetikzlibrary{arrows,calc,patterns}
\tikzstyle{Rect}=[draw=blue,line width=0.001pt,preaction={clip, postaction={pattern=north east lines, pattern color=blue,line width=0.1pt}}]
\tikzset{
	>=stealth',
	help lines/.style={dashed, thick},
	axis/.style={<->},
	important line/.style={thick},
	connection/.style={thick, dotted},
}

\newtheorem{lemma}{Lemma}[section]
\newtheorem{example}{Example}[section]

\usepackage{eso-pic}
\AddToShipoutPictureBG*{%
	\AtPageUpperLeft{%
		\setlength\unitlength{1in}%
		\hspace*{\dimexpr0.5\paperwidth\relax}
		\makebox(0,-1.75)[c]{
			\begin{tabular}{c c}
				Rodrigo A. Gonz\'alez et al.,
				Parsimonious Identification of Continuous-Time Systems: A Block-Coordinate Descent Approach. \\
				To appear in
				{\em 22nd IFAC World Congress},
				Yokohama, Japan, 2023
				uploaded to ArXiv on \today \\
\end{tabular}}}}

\begin{document}
\begin{frontmatter}

\title{Parsimonious Identification of Continuous-Time Systems: A Block-Coordinate Descent Approach\thanksref{footnoteinfo}}

\thanks[footnoteinfo]{This work was supported by the Swedish Research Council under contract number 2016-06079 (NewLEADS), by the Digital Futures project EXTREMUM, and by the research program VIDI with project number 15698, which is (partly) financed by the Netherlands Organization for Scientific Research (NWO).}

\author[First]{Rodrigo A. Gonz\'alez} 
\author[Second]{Cristian R. Rojas} 
\author[Third]{Siqi Pan}
\author[Third]{James S. Welsh}

\address[First]{Department of Mechanical Engineering, Eindhoven University of Technology, Eindhoven, The Netherlands}
\address[Second]{Division of Decision and Control Systems, KTH Royal Institute of Technology, Stockholm, Sweden}
\address[Third]{School of Engineering, University of Newcastle, Callaghan, NSW, Australia}

\begin{abstract}
	The identification of electrical, mechanical, and biological systems using data can benefit greatly from prior knowledge extracted from physical modeling. Parametric continuous-time identification methods can naturally incorporate this knowledge, which leads to interpretable and parsimonious models. However, some applications lead to model structures that lack parsimonious descriptions using unfactored transfer functions, which are commonly used in standard direct approaches for continuous-time system identification. In this paper we characterize this parsimony problem, and develop a block-coordinate descent algorithm that delivers parsimonious models by sequentially estimating an additive decomposition of the transfer function of interest. Numerical simulations show the efficacy of the proposed approach.
\end{abstract}


\end{frontmatter}

\section{Introduction}
\vspace{-0.2cm}Continuous-time system identification methods \citep{Garnier2008book} are popular and successful in a wide range of practical applications thanks to several advantages they enjoy compared to the discrete-time algorithms \citep{soderstrom1989system}. One of these traits is that some continuous-time methods allow the direct incorporation of the \textit{a priori} knowledge of the relative degree of the physical systems they model. This leads to more parsimonious representations, which means that simpler models (in terms of number of parameters) can be used to accurately describe the phenomenon at hand.

With regards to model flexibility, most linear continuous-time identification methods parameterize the model structure as an unfactored transfer function with a user-defined number of poles and zeros. This is the case for the Prediction Error Method for continuous-time systems (PEM), as well as the Simplified Refined Instrumental Variable method for Continuous-time systems (SRIVC, \cite{young1980refined}) and the Least-Squares State-Variable Filter method (LSSVF, \cite{young1965determination}). However, some practical applications related to, e.g., motion systems and vibration analysis, consider systems that are more easily interpreted as a sum of transfer functions with distinct denominators, typically corresponding to different resonant modes. The SRIVC method cannot handle the estimation of such additive continuous-time systems, since such model structure does not yield a pseudolinear regression suitable for constructing the filtered instrument and regressor vectors \citep{garnier2007optimal}. In addition, it is known that numerical conditioning issues may arise when estimating high-order or highly-resonant systems, which are typically parameterized as the sum of second-order continuous-time systems \citep{gilson2017frequency}.

In this paper we propose an algorithm that delivers parsimonious models for additive continuous-time systems. This methods performs a block coordinate descent using SRIVC as a tool for decreasing the output error cost at each iteration. More precisely,
\begin{enumerate}[label=C\arabic*]
	\item
	We obtain an explicit condition under which it is preferable to estimate additive models instead of unfactored transfer functions, which is relevant when deciding what model structure should be considered for continuous-time system identification;
	
	\item
	We prove the global convergence (for large sample size) of a coordinate-block descent algorithm for the identification of additive continuous-time systems;
	
	\item 
	We provide closed-form expressions of the SRIVC iterations that, at convergence for finite sample size and under mild conditions, are proven to deliver a critical point of the cost function being minimized at each block-coordinate descent step.
\end{enumerate}
 
The paper is structured as follows. In Section \ref{sec:system}, we introduce the problem setup and discuss different model parameterizations. Section \ref{sec:parsimony} introduces the parsimony problem for continuous-time models, while the block-coordinate descent method that solves this problem is presented in Section \ref{sec:block}. A numerical study can be found in Section \ref{sec:simulations}, and the paper is concluded in Section \ref{sec:conclusions}. Proofs of the main results can be found in the Appendix.

\vspace{-0.115cm}
\section{System and model setup}
\vspace{-0.15cm}\label{sec:system}
Consider the single-input single-output (SISO), linear and time-invariant, continuous-time system
\begin{equation}
	\label{ctsystem1}
	x(t) =  \frac{B^*(p)}{A^*(p)} u(t),
\end{equation}
where $p$ is the Heaviside operator, and $u(t)$ is the input signal. The numerator and denominator polynomials $B^*(p)$ and $A^*(p)$ are assumed coprime and given by
\begin{align}
	A^*(p)&= a_n^*p^n+a_{n-1}^*p^{n-1}+\cdots + a_1^*p + 1, \notag \\
	B^*(p)&= b_m^*p^{m}+b_{m-1}^*p^{m-1}+\cdots + b_1^*p + b_0^*, \notag
\end{align}
with $a_n^*\neq 0$ and $n\geq m$. The polynomials $A^*(p)$ and $B^*(p)$ are jointly described by the parameter vector 
\begin{equation}
	\label{ctparametervector}
	\bm{\theta}^* = \begin{bmatrix}
		a_1^*, & a_2^*, & \dots, & a_n^*, & b_0^*, & b_1^*, & \dots, & b_m^*
	\end{bmatrix}^\top.
\end{equation}
The system $G^*\hspace{-0.02cm}(p)=B^*\hspace{-0.02cm}(p)/A^*\hspace{-0.02cm}(p)$ can also be described in its modal (or \textit{additive}) form. This modal form, which describes the system as a finite sum of transfer functions \citep{pan2021continuous}, leads to the alternative description
\begin{equation}
	\label{system1}
	x(t) = \sum_{i=1}^K G_i^*(p) u(t),
\end{equation}
where $G_i^*(p) = B_i^*(p)/A_i^*(p)$, and the transfer function polynomials $A_i^*(p)$ and $B_i^*(p)$ have degrees $n_i$ and $m_i$, respectively ($m_i\leq n_i$). We assume without loss of generality that the $A_i^*(p)$ polynomials are anti-monic (i.e., their constant coefficient is fixed to 1), and they are jointly coprime. For this representation we write $\bm{\theta}_i^*$ as the parameter vector that describes the transfer function $G^*_i(p)$, similarly to \eqref{ctparametervector}. In addition, for each separate submodel to be identifiable, we assume that at most one subsystem $G_i^*(p)$ is biproper.

Given the system in \eqref{system1}, a noisy output measurement is retrieved every $h$[s]. That is,
\begin{equation}
	y(kh) = x(kh)+v(kh), \notag
\end{equation}
where $v(kh)$ is assumed to be a zero-mean stationary random process of variance $\sigma^2$ that is uncorrelated with the input. We assume that the input has a zero-order hold (ZOH) intersample behavior, although our results can also be extended to arbitrary inputs \citep{gonzalez2020consistent}.

This paper studies how to determine a model for the additive decomposition of $G^*(p)$ in \eqref{system1}, with known model structure, based on $N$ input and output data samples $\{u(kh),y(kh)\}_{k=1}^N$. In addition, we are interested in comparing both standard \eqref{ctsystem1} and additive \eqref{system1} forms in terms of their parsimony when implementing continuous-time system identification methods. As the models depend on the parameter vector $\bm{\theta}$ that is being estimated, we write them as $G(p,\bm{\theta})$, or $G(p)$, if the dependence in $\bm{\theta}$ is obvious.

\vspace{-0.1cm}
\section{Parsimony in continuous-time system identification}
\label{sec:parsimony}
\vspace{-0.1cm}
Direct continuous-time methods such as the SRIVC estimator may suffer from a lack of parsimony when we identify the sum of transfer functions of particular relative degrees. The following proposition, which constitutes Contribution C1 of this paper, indicates the number of additional parameters that are being estimated if the user decides to estimate \eqref{system1} with a model structure $G(p)=B(p)/A(p)$ of relative degree $r$ instead of estimating the parameters of each transfer function $B_i^*(p)/A_i^*(p)$ separately.

\begin{prop}
	\label{prop1}
	Consider the system in \eqref{system1}, and the model structure $G(p)=\sum_{i=1}^K B_i(p)/A_i(p)$. If one instead decides to use the model structure $G(p)=B(p)/A(p)$ for identification, with minimal relative degree $r$ that contains the true system, then the latter model structure incurs in a lack of parsimony if and only if
	\begin{equation}
		\label{excess}
		\sum_{i=1}^{K} r_i - r> K-1,
	\end{equation}
	where $r_i=n_i-m_i$ is the relative degree of $B_i(p)/A_i(p)$. The excess in \eqref{excess}, i.e., the difference between the left and right hand sides, is the number of additional parameters that the model structure $G(p)=B(p)/A(p)$ considers. 
\end{prop}

\begin{pf}
The number of parameters to be estimated in each model $B_i(p)/A_i(p)$ is $n_i+m_i+1$, which leads to the need of estimating $\sum_{i=1}^K (n_i+m_i) + K$ parameters if one considers the model structure $G(p)=\sum_{i=1}^K B_i(p)/A_i(p)$. On the other hand, the model structure $G(p)=B(p)/A(p)$ requires estimating $2\sum_{i=1}^K n_i - r +1$ parameters. Subtracting both of these expressions leads to an excess of parameters given by $\sum_{i=1}^{K} r_i - r - K+1$. If such quantity is greater than zero, then we reach the condition in \eqref{excess}.  \hspace{1.8cm} \hfill \qed 
\end{pf}

A consequence of this result is that if $r=0$ or $r=1$, then the model structure $G(p)=B(p)/A(p)$ suffers from a lack of parsimony if there exists a transfer function $G_i^*(p)$ in the expansion \eqref{system1} with relative degree greater than one. 
\begin{example}
	Consider the system
	\begin{equation}
		\label{examplesystem}
		G^*(p) = \frac{3}{0.25p^2 + 0.25 p +1}+\frac{1}{0.0025p^2 +0.01 p +1}.
	\end{equation}
	This system corresponds to a truncated modal description of a flexible structure, such as a piezoelectric laminate beam \citep{moheimani2003spatial}. Only 6 parameters must be estimated if the following model structure is used:
	\begin{equation}
		\label{parsimoniousmodelstructure}
		G(p)=\frac{b_{01}}{a_{21} p^2 +a_{11} p +1} + \frac{b_{02}}{a_{22} p^2 +a_{12} p +1}.
	\end{equation}
	On the other hand, if this modal decomposition is not taken into account and one decides to estimate the model
	\begin{equation}
		\label{parameterization2}
		G(p)=\frac{b_{2}p^2+b_{1}p+b_{0}}{a_4 p^4 + a_{3} p^3 + a_{2} p^2 + a_{1} p +1} 
	\end{equation}
	with no constraints on the parameter values, then 7 parameters must be estimated. This model structure leads to a lack of parsimony compared to \eqref{parsimoniousmodelstructure}.
\end{example}
%
\vspace{-0.1cm}
\section{Block-coordinate descent method for continuous-time system identification}
\label{sec:block}
\vspace{-0.1cm}
In this section we present a method that solves the parsimony problem described in Section \ref{sec:parsimony} for the identification of linear continuous-time systems. The goal is to estimate the parameters of additive models of the form \eqref{system1} by solving the following minimization problem:
\begin{align}
	\label{cost}
	\min_{\substack{\bm{\theta}_i \in \Omega_i,\\ i = 1, \dots, K}} \frac{1}{N} \sum_{k=1}^N \left[ y(kh) - \sum_{i=1}^K G_i(p,\bm{\theta}_i) u(kh) \right]^2,
\end{align}
with $\Omega_i\subset \mathbb{R}^{n_i+m_i+1}$ being a compact set where the parameters of the $i$th subsystem are assumed to lie. Note that to solve the optimization problem in \eqref{cost} one cannot directly apply refined instrumental variable methods (i.e., the SRIVC method \citep{young1980refined}), since the denominator polynomials of each submodel $G_i(p,\bm{\theta}_i)$ are distinct. Instead, we propose a \emph{block-coordinate descent} algorithm, in which the cost function is iteratively minimized with respect to $\bm{\theta}_i$ while leaving the other decision variables fixed. To this end, we define the total cost function
\begin{align}
	V_{\hspace{-0.02cm}N}\hspace{-0.02cm}(\bm{\theta}_1, \hspace{-0.02cm}\dots\hspace{-0.02cm}, \hspace{-0.02cm}\bm{\theta}_{\hspace{-0.02cm}K}\hspace{-0.02cm})\hspace{-0.09cm} := \hspace{-0.09cm}\frac{1}{N} \hspace{-0.06cm}\sum_{k=1}^N \hspace{-0.04cm}\left[ \hspace{-0.02cm}y(kh) \hspace{-0.06cm}-\hspace{-0.07cm} \sum_{i=1}^K \hspace{-0.05cm} G_i\hspace{-0.02cm}(p,\hspace{-0.02cm}\bm{\theta}_i) u(kh) \hspace{-0.04cm} \right]^{\hspace{-0.03cm}2}\hspace{-0.07cm},\hspace{-0.05cm} \notag
\end{align}
where $\bm{\theta}_i \in \Omega_i$ for $i = 1, \dots, K$. Algorithm \ref{algorithm1} describes the general proposed procedure.

\begin{algorithm}
	\caption{Block-coordinate descent algorithm}
	\begin{algorithmic}[1]
		\State Input: initial parameter vector $\bm{\theta}_i^1$ for each $i$
		\For{$l = 1, 2, \dots$}
		\For{$i = 1, \dots, K$}
		\State $\hspace{-0.15cm}\bm{\theta}_i^{l\hspace{-0.01cm}+\hspace{-0.01cm}1} \hspace{-0.12cm}\gets \hspace{-0.06cm} \arg\hspace{-0.02cm}\min\limits_{\hspace{-0.53cm}\bm{\theta}_i \in \Omega_i} V\hspace{-0.04cm}(\bm{\theta}_1^{l+1}\hspace{-0.02cm},\hspace{-0.02cm} \dots\hspace{-0.02cm}, \bm{\theta}_{i-1}^{l+1}\hspace{-0.02cm},\hspace{-0.02cm} \bm{\theta}_i\hspace{-0.02cm},\hspace{-0.02cm} \bm{\theta}_{i+1}^{l}\hspace{-0.02cm},\hspace{-0.02cm} \dots\hspace{-0.02cm},\hspace{-0.02cm} \bm{\theta}_K^{l}\hspace{-0.02cm})$
		\EndFor
		\EndFor
		\State Output: parameter vectors $\lim_{l\to \infty} \bm{\theta}_i^{l}$, $i=1,2,\dots,K$.
	\end{algorithmic}
	\label{algorithm1}
\end{algorithm}

Let $\mathbf{x}^l := [(\bm{\theta}_1^l)^\top, \; \dots, \; (\bm{\theta}_K^l)^\top]^\top \in \Omega \subset \mathbb{R}^P$, with $\Omega=\prod_{i=1}^{K} \Omega_i$ being the parameter space, and $P=\sum_{i=1}^K (n_i+m_i) + K$ being the total number of parameters to estimate. Each iteration of Algorithm \ref{algorithm1} (in $l$) can therefore be written as $\mathbf{x}^{l+1} = \mathcal{A}(\mathbf{x}^l)$, where $\mathcal{A}\colon \mathbb{R}^P \to \mathbb{R}^P$ is a mapping that can be described as a composition of functions:
\begin{equation}
	\label{algorithma}
	\mathcal{A} = \mathbf{S} \circ \mathbf{C}^K \circ \mathbf{S} \circ \mathbf{C}^{K-1} \circ \cdots \circ \mathbf{S} \circ \mathbf{C}^1,
\end{equation}
where we denote the choice function $\mathbf{C}^i(\mathbf{x}) := (\mathbf{x}, i)$, the joint parameter vector $\mathbf{x} := [\bm{\theta}_1^\top, \dots , \bm{\theta}_K^\top]^\top$, and the optimization step $\mathbf{S}(\mathbf{x},i) := (\bm{\theta}_1, \dots,\overline{\bm{\theta}}_i,\dots, \bm{\theta}_K)$, with $\overline{\bm{\theta}}_i = \arg\min_{\bm{\theta}_i \in \Omega_i} V_N(\bm{\theta}_1, \dots, \bm{\theta}_i, \dots, \bm{\theta}_K)$. The following result concerns the convergence of Algorithm \ref{algorithm1} to a stationary point of the cost \eqref{cost}, and constitutes Contribution C2.

\begin{thm}[Global convergence of Algorithm \ref{algorithm1}]
	\label{thm42}
	Consi-\textcolor{white}{space} der the set of fixed points $\Gamma_N = \{\mathbf{x}\in \Omega\colon \nabla V_N(\mathbf{x}) = \mathbf{0}\}$. For a sufficiently large $N$, the limit of any convergent subsequence of $\{\mathbf{x}^l\}$ obtained from the iterations $\mathbf{x}^{l+1} = \mathcal{A}(\mathbf{x}^l)$ almost surely belongs to $\Gamma_N$.
\end{thm}
\begin{pf}
	See Appendix \ref{appendixthm42}. \hspace{4.2cm} \hfill \qed
\end{pf}

A crucial part of Theorem \ref{thm42} is the characterization of the solution set $\Gamma_N$. One can establish that, in addition to the uniform convergence of $V_N$ as $N$ tends to infinity, the gradient $\nabla V_N$ also converges uniformly to $\nabla \bar{V}$, where $\bar{V}$ is defined in Eq. \eqref{barv}. This implies that the set of stationary points of $V_N$, $\Gamma_N$, converges to those of $\bar{V}$, say, $\bar{\Gamma} = \{ \mathbf{x} \in \Omega\colon \nabla \bar{V}(\mathbf{x}) = \mathbf{0} \}$, in the sense that for every $\epsilon > 0$ and almost every realization $\xi$, there exists a sample size $N(\xi, \epsilon) \in \mathbb{N}$ such that for all $N \geq N(\xi, \epsilon)$ and every $\mathbf{x} \in \Gamma_N$ there is a $\mathbf{y} \in \bar{\Gamma}$ such that $\| \mathbf{x} - \mathbf{y} \| < \epsilon$, and also for every $\mathbf{y} \in \bar{\Gamma}$ there is a $\mathbf{x} \in \Gamma_N$ such that $\| \mathbf{x} - \mathbf{y} \| < \epsilon$.

Unfortunately, $\bar{\Gamma}$ may contain not only the global minima of $\bar{V}$ (corresponding to all permutations $\bm{\theta}_i$, $\bm{\theta}_j,i\neq j$, that share the same model structure), but also other local minima and saddle points \cite[Section~7.1]{regalia1995adaptive}. Some references give sufficient conditions under which $\bar{\Gamma}$ contains only the global minima. For example, if $\{u(kh)\}_{k \in \mathbb{Z}}$ is white noise and the model structure exactly contains the true system, then according to Lemma 4.1 of \cite{soderstrom1975uniqueness}, every stationary point which is not a global minimizer of $\bar{f}$ must give pole-zero cancellations, thereby leading to the same minimal realization. This result, however, is not valid in general for arbitrary inputs.

\begin{remark}
	\label{remarkglobal}
	For Theorem \ref{thm42} to hold, it is only required for $V_N$ to be decreasing at each step. This means that it is sufficient to find a parameter vector $\bar{\bm{\theta}}_i$ that reduces the cost $V_N$ instead of minimizing it. 
\end{remark}

\subsection{SRIVC for computing the descent step}

The block-coordinate descent algorithm described in Algorithm \ref{algorithm1} requires a way to compute $\mathbf{S}(\mathbf{x},i)$ at each iteration, for each $i=1,2,\dots,K$. That is, we need to compute
\begin{align}
	\bm{\theta}_i^{l+1} &= \underset{\bm{\theta}_i \in \Omega_i}{\arg \min} \frac{1}{N} \sum_{k=1}^N \bigg[ y(kh) \hspace{-0.08cm}-\hspace{-0.07cm} \sum_{j=1}^{i-1} G_j(p,\bm{\theta}_j^{l+1}) u(kh) \notag \\
	\label{opt}
	&-\hspace{-0.12cm}\sum_{j=i+1}^{K} \hspace{-0.05cm}G_j(p,\bm{\theta}_j^l) u(kh) - G_i(p,\bm{\theta}_i) u(kh) \bigg]^2
\end{align}
for $i=1,2,\dots, K$. The key insight is that, for $\Omega_i=\mathbb{R}^{n_i+m_i+1}$ and fixed values of $\{\bm{\theta}_j^{l+1}\}_{j=1}^{i-1}, \{\bm{\theta}_j^l\}_{j=i+1}^{K}$, the optimization problem in \eqref{opt} reduces to a nonlinear least-squares problem that can be solved via SRIVC iterations. Indeed, if we define the residual output of each submodel
\begin{equation}
	\tilde{y}(kh) \hspace{-0.09cm}:=\hspace{-0.03cm} y(kh)\hspace{-0.01cm}-\hspace{-0.02cm} \sum_{j=1}^{i-1} \hspace{-0.06cm}G_{\hspace{-0.02cm}j}\hspace{-0.03cm}(p,\hspace{-0.02cm}\bm{\theta}_j^{l\hspace{-0.02cm}+\hspace{-0.02cm}1}\hspace{-0.02cm}) u(kh)\hspace{-0.01cm}-\hspace{-0.1cm}\sum_{j=i+1}^{K} \hspace{-0.15cm}G_{\hspace{-0.02cm}j}\hspace{-0.02cm}(p,\hspace{-0.02cm}\bm{\theta}_j^l) u(kh), \notag
\end{equation}
then $\bm{\theta}_i^{l+1}$ must satisfy the first-order optimality condition
\begin{equation}
	\label{convergingto}
	\frac{1}{N}\sum_{k=1}^N \hat{\bm{\varphi}}_\textnormal{f}(kh,\bm{\theta}_i^{l+1}) e(kh,\bm{\theta}_i^{l+1}) = \mathbf{0},
\end{equation}
where the gradient and total residual are, respectively,
\begin{align}
	&\hspace{-0.23cm}\hat{\bm{\varphi}}_\textnormal{f}(kh,\hspace{-0.02cm}\bm{\theta}_i^{l\hspace{-0.02cm}+\hspace{-0.02cm}1}\hspace{-0.02cm}) \hspace{-0.05cm}=\hspace{-0.06cm} \bigg[\hspace{-0.02cm}\frac{-p \hspace{-0.02cm}B_i^{l\hspace{-0.02cm}+\hspace{-0.02cm}1}(p)}{[A_i^{l\hspace{-0.02cm}+\hspace{-0.02cm}1}(p)]^2}\hspace{-0.02cm} u(kh), \dots, \hspace{-0.05cm}\frac{-p^n \hspace{-0.05cm}B_i^{l\hspace{-0.02cm}+\hspace{-0.02cm}1}(p)}{[A_i^{l\hspace{-0.02cm}+\hspace{-0.02cm}1}(p)]^2} u(kh), \notag \\
	&\hspace{2.1cm} \frac{1}{A_i^{l\hspace{-0.02cm}+\hspace{-0.02cm}1}(p)}u(kh),\dots, \hspace{-0.03cm} \frac{p^m}{A_i^{l\hspace{-0.02cm}+\hspace{-0.02cm}1}(p)} u(kh) \hspace{-0.02cm}\bigg]^{\hspace{-0.02cm}\top}\hspace{-0.03cm}, \hspace{-0.1cm} \notag  \\
	&\hspace{-0.06cm}e(kh,\bm{\theta}_i^{l\hspace{-0.02cm}+\hspace{-0.02cm}1}) \hspace{-0.04cm}=\hspace{-0.02cm} \tilde{y}(kh)- \frac{B_i^{l\hspace{-0.02cm}+\hspace{-0.02cm}1}(p)}{A_i^{l\hspace{-0.02cm}+\hspace{-0.02cm}1}(p)}u(kh), \notag
\end{align}
with $B_i^{l+1}$ and $A_i^{l+1}$ being the numerator and denominator polynomials of the $i$th submodel evaluated at $\bm{\theta}_{i}^{l+1}$.

Lemma \ref{lemmagnsrivc}, which constitutes Contribution C3 of this paper, provides the SRIVC iterations that are shown to deliver stationary points of the cost in \eqref{opt} at convergence in iterations under mild conditions.
\begin{lemma}[SRIVC iterations]
	\label{lemmagnsrivc}
	For an initial mo\-del parameter estimate $\bm{\theta}_{i,0}^{l+1}$ and $s=0,1,2,\dots$, consider the following SRIVC iterations
	\begin{subequations}
		\label{formulasrivc}
		\begin{align}
			\label{matrixinversesrivc}
			\bm{\theta}_{i,s+1}^{l+1} &= \left[\frac{1}{N}\sum_{k=1}^{N}\hat{\bm{\varphi}}_\textnormal{f}(kh,\bm{\theta}_{i,s}^{l+1}) \bm{\varphi}_\textnormal{f}^\top(kh,\bm{\theta}_{i,s}^{l+1})\right]^{-1} \\
			&\hspace{0.5cm}\times \hspace{-0.04cm} \left[\hspace{-0.04cm}\frac{1}{N}\hspace{-0.04cm}\sum_{k=1}^{N}\hat{\bm{\varphi}}_\textnormal{f}(kh,\bm{\theta}_{i,s}^{l+1}) \tilde{y}_{\textnormal{f}}(kh,\bm{\theta}_{i,s}^{l+1})\right]\hspace{-0.02cm}, \notag
		\end{align}
	\end{subequations}
	where the filtered regressor $\bm{\varphi}_\textnormal{f}$ and filtered residual output $\tilde{y}_{\textnormal{f}}$ are given by
	\begin{align}
		&\hspace{-0.2cm}\bm{\varphi}_\textnormal{f}(kh,\bm{\theta}_{i,s}^{l\hspace{-0.02cm}+\hspace{-0.02cm}1}) \hspace{-0.05cm}=\hspace{-0.06cm}  \bigg[\frac{-p}{A_{i,s}^{l\hspace{-0.02cm}+\hspace{-0.02cm}1}(p)} \tilde{y}(kh), \dots, \hspace{-0.03cm}\frac{-p^n}{A_{i,s}^{l\hspace{-0.02cm}+\hspace{-0.02cm}1}(p)} \tilde{y}(kh), \notag \\
		\label{filteredregressor}
		&\hspace{2cm} \frac{1}{A_{i,s}^{l\hspace{-0.02cm}+\hspace{-0.02cm}1}(p)}u(kh),\dots, \hspace{-0.03cm} \frac{p^m}{A_{i,s}^{l\hspace{-0.02cm}+\hspace{-0.02cm}1}(p)} u(kh) \hspace{-0.02cm}\bigg]^{\hspace{-0.04cm}\top}\hspace{-0.03cm}, \hspace{-0.1cm} \\
		\label{filteredoutput}
		&\hspace{-0.1cm}\tilde{y}_\textnormal{f}(kh,\bm{\theta}_{i,s}^{l\hspace{-0.02cm}+\hspace{-0.02cm}1}) \hspace{-0.05cm}=\hspace{-0.06cm} \frac{1}{A_{i,s}^{l\hspace{-0.02cm}+\hspace{-0.02cm}1}(p)}\tilde{y}(kh).
	\end{align}
	If the matrix being inverted in \eqref{matrixinversesrivc} is non-singular for all integers $s$ large enough, then any converging point (when $s$ tends to infinity), if they exist, satisfies the first-order optimality condition \eqref{convergingto}.
\end{lemma}
\begin{pf}
	See Appendix \ref{prooflemmagn}. \hspace{4.2cm} \hfill \qed
\end{pf}

In practice, one can terminate the SRIVC procedure for each submodel whenever $V_N$ has strictly decreased from its initial value. This termination rule is in agreement with Remark \ref{remarkglobal}. Also, note that a decrease in the cost function requires to initialize the methods close to the global optimum, since the SRIVC method does not guarantee global convergence for finite-sample size. This can be done by applying the standard SRIVC estimator to find a model for \eqref{ctsystem1}, finding the partial fraction expansion, and then deleting the unwanted high-order numerator terms.

\begin{remark}
	The non-singularity of the matrix in \eqref{matrixinversesrivc} depends on the persistence of excitation of the input, as well as on the interpolation error when constructing the filtered output in the regressor vector. The generic non-singularity result in Theorem 1 of \cite{pan2020consistency} can be extended to the case in \eqref{matrixinversesrivc} by including the model parameters of the other submodels as part of the genericity statement. Such proof, however, is outside of the scope of the current paper.
\end{remark}

The method we propose for identifying additive continuous-time systems is detailed in Algorithm \ref{completealgorithm}. Apart from what has been discussed, additional techniques can be fit to the algorithm to robustify it. These techniques include a) including a non-fixed step size \citep{soderstrom1982some} in the incremental form of these iterations, b) admitting unstable models by ad-hoc prefiltering \citep{gonzalez2022unstable}, and c) introducing randomization when choosing the next submodel to be updated (i.e., the $i$ index). The details of these extensions are left out of our exposition for simplicity only.

\begin{algorithm}
	\renewcommand{\thealgorithm}{2}
	\caption{\hspace{-0.1cm}: Block-coordinate descent method for additive continuous-time system identification}
	\begin{algorithmic}[1]
		\State Input:$\hspace{0.03cm}\{\hspace{-0.03cm}u(\hspace{-0.023cm}kh\hspace{-0.023cm}),\hspace{-0.025cm}y(\hspace{-0.023cm}kh\hspace{-0.023cm})\hspace{-0.023cm}\}_{\hspace{-0.03cm}k\hspace{-0.01cm}=\hspace{-0.02cm}1}^{\hspace{-0.03cm}N}\hspace{-0.04cm},\hspace{0.04cm}$initialization $\mathbf{x}_{\hspace{-0.015cm}1}\hspace{-0.13cm}=\hspace{-0.12cm}[\bm{\theta}_{1}^{\hspace{-0.02cm}1\hspace{-0.04cm}\top}\hspace{-0.06cm},\hspace{-0.02cm}.\hspace{0.03cm}.\hspace{0.03cm}.\hspace{0.03cm},\hspace{-0.02cm}\bm{\theta}_{K}^{\hspace{-0.02cm}1\hspace{-0.04cm}\top}]^{\hspace{-0.05cm}\top}\hspace{-0.08cm}$, tolerance factor $\epsilon$, maximum number of $\mathcal{A}$-iterations $M$, and maximum number of SRIVC iterations $M_\textnormal{s}$ 
		\State $l\gets 1$, $s\gets 1$,
		\While{$l\leq M$}
		\For{$i = 1, \dots, K$}
		\State $\bm{\theta}_{i,1}^{l+1}\gets \bm{\theta}_{i}^l$, $s\gets 1$
		\While{$s\leq M_\textnormal{s}$}
		\State Compute $\bm{\theta}_{i,s+1}^{l+1}$ using SRIVC \eqref{formulasrivc}
		\If{\hspace{-0.06cm}$V\hspace{-0.04cm}(\hspace{-0.02cm}\bm{\theta}_1^{l\hspace{-0.02cm}+\hspace{-0.02cm}1}\hspace{-0.07cm}, \hspace{-0.03cm} .\hspace{0.03cm}.\hspace{0.03cm}., \hspace{-0.03cm}\bm{\theta}_{\hspace{-0.02cm}i\hspace{-0.02cm}-\hspace{-0.02cm}1}^{l\hspace{-0.02cm}+\hspace{-0.02cm}1}\hspace{-0.03cm},\hspace{-0.03cm} \bm{\theta}_{\hspace{-0.02cm}i,\hspace{-0.01cm}s\hspace{-0.02cm}+\hspace{-0.02cm}1}^{l\hspace{-0.02cm}+\hspace{-0.02cm}1}\hspace{-0.03cm},\hspace{-0.04cm} \bm{\theta}_{\hspace{-0.02cm}i\hspace{-0.02cm}+\hspace{-0.02cm}1}^{l}\hspace{-0.03cm}, \hspace{-0.03cm} .\hspace{0.03cm}.\hspace{0.03cm}.,\hspace{-0.03cm} \bm{\theta}_{\hspace{-0.05cm}K}^{l}\hspace{-0.04cm})\hspace{-0.1cm}<\hspace{-0.1cm}V\hspace{-0.04cm}(\hspace{-0.02cm}\bm{\theta}_1^{l\hspace{-0.02cm}+\hspace{-0.02cm}1}\hspace{-0.08cm},$
			\hspace*{1.75cm} $.\hspace{0.03cm}.\hspace{0.03cm}.,\bm{\theta}_{i\hspace{-0.02cm}-\hspace{-0.02cm}1}^{l\hspace{-0.02cm}+\hspace{-0.02cm}1},\hspace{-0.02cm} \bm{\theta}_{i,1}^{l\hspace{-0.02cm}+\hspace{-0.02cm}1}, \bm{\theta}_{i\hspace{-0.02cm}+\hspace{-0.02cm}1}^{l}, \hspace{-0.03cm} .\hspace{0.03cm}.\hspace{0.03cm}., \bm{\theta}_{\hspace{-0.05cm}K}^{l}\hspace{-0.04cm})$}
		\State $\bm{\theta}_{i}^{l+1}\gets \bm{\theta}_{i,s+1}^{l+1}$, $s\gets M_\textnormal{s}$
		\EndIf
		\State $s\gets s+1$
		\EndWhile{}
		\EndFor
		\State $\mathbf{x}_{l+1}\gets [\bm{\theta}_{1}^{l+1 \top},\dots,\bm{\theta}_{K}^{l+1\top}]^\top$
		\If{$\dfrac{\|\mathbf{x}_{l+1}-\mathbf{x}_l\|_2}{\|\mathbf{x}_l\|_2}<\epsilon$}
		\State $\hat{\mathbf{x}}\gets \mathbf{x}_{l+1}$, $l \gets M$
		\EndIf
		\State $l \gets l+1$
		\EndWhile{}
		\State Output: $\hat{\mathbf{x}}$ and its associated models $\{\hat{G}_i(p)\}_{i=1}^K$.
	\end{algorithmic}
	\label{completealgorithm}
\end{algorithm}
\vspace{-0.1cm}
\section{Simulations}
\label{sec:simulations}
\vspace{-0.15cm}
We now verify the applicability of the proposed method through two numerical experiments. 
\vspace{-0.15cm}
\subsection{Case Study 1: 4th order system}
\vspace{-0.15cm}
We consider $G^*(p)$ as in \eqref{examplesystem}, which can also be written as
\begin{equation}
	\label{oftheform}
	G^*\hspace{-0.03cm}(p) \hspace{-0.07cm}=\hspace{-0.07cm} \frac{0.2575p^2+0.28p+4}{0.000625p^4\hspace{-0.08cm}+\hspace{-0.06cm}0.003125p^3\hspace{-0.08cm}+\hspace{-0.06cm}0.255p^2\hspace{-0.08cm}+\hspace{-0.06cm}0.26p\hspace{-0.08cm}+\hspace{-0.08cm}1}. \hspace{-0.1cm}
\end{equation}
Note that the denominator polynomial coincides with that of the Rao-Garnier system, which is a benchmark for linear continuous-time system identification  \citep{rao2002numerical}. We compare the proposed block-coordinate descent (BCD) method with the SRIVC estimator and the indirect approach using the SRIV estimator \citep{young1976some} converted to continuous-time. The standard SRIVC estimator uses $n=4$ and $m=2$ as the model polynomial degrees, in agreement with the model structure \eqref{parameterization2}. A Monte Carlo simulation is performed in order to test the fit and the mean square error (MSE) of the estimated parameters of all methods. The input is a Gaussian white noise of unit variance that is interpolated with a ZOH, and the measurement noise is also white and Gaussian, of unitary variance. Five hundred Monte Carlo runs are computed with $N=10000$ and $h=0.005 [\textnormal{s}]$. In each run, all methods are initialized with an additive model whose parameters deviate at most $10\%$ from the true parameters.

Once the model is obtained from the data in each Monte Carlo run, the fit metric is computed by
\begin{equation}
	\textnormal{fit} = 100 \left(1- \|\hat{\mathbf{x}}^i-\mathbf{x}\|_2/\|\mathbf{x}-\bar{x}\mathbf{1}\|_2 \right), \notag
\end{equation}
where $\mathbf{x}$ is the noiseless output sequence, $\hat{\mathbf{x}}^i$ is the simulated output sequence of the $i$th estimated model, and $\bar{x}$ is the average value of $\{x(kh)\}_{k=1}^N$.

\begin{figure}
	\centering{
		\includegraphics[width=0.4\textwidth]{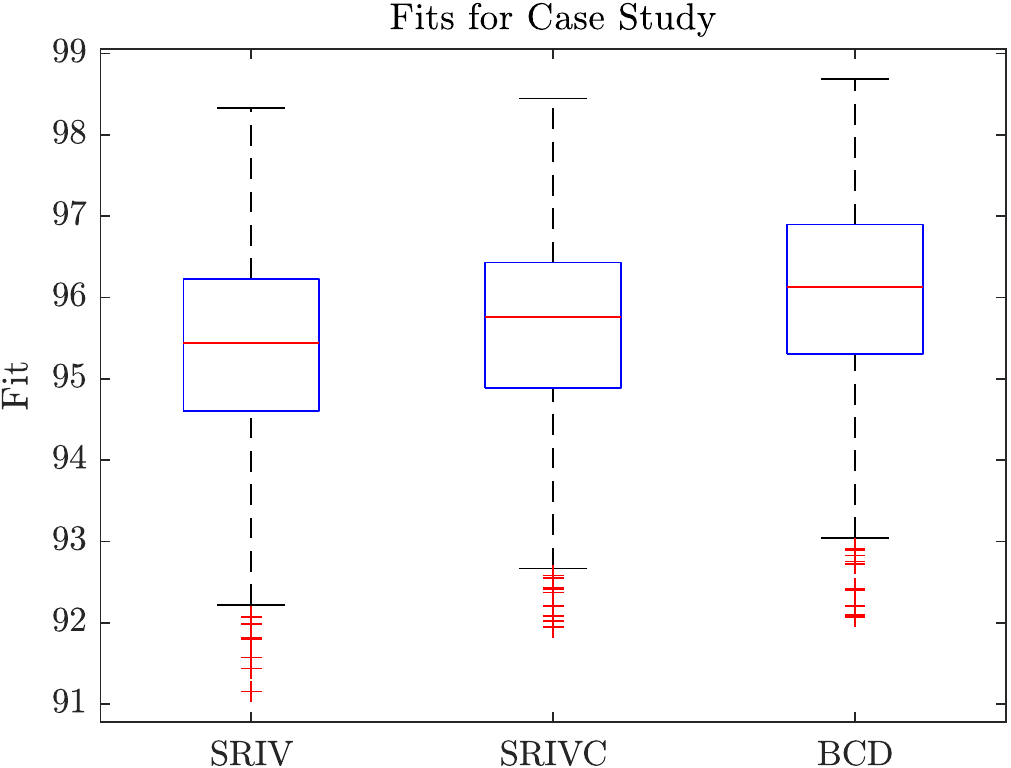}
		\caption{Boxplot of the fit metric for Case Study 1.}
		\label{fig1}}
\end{figure}

In Figure \ref{fig1} we observe the benefit of obtaining parsimonious models by plotting the boxplot for the fit metric of all methods. We have also computed the MSE of each parameter of the equivalent sum model of the form \eqref{oftheform} in Table \ref{table1}. The indirect approach method (SRIV) provides models with relative degree equal to one almost always, hence inducing over-parametrization. The SRIVC estimator avoids this over-parametrization issue yet still cannot provide parsimonious models due to Proposition \ref{prop1}. The BCD method is the only one that provides the correct model structure, which leads to the best performance in the fit metric and MSEs of each estimated parameter.

\begin{table*}
	\centering
	\caption{MSEs of the estimated parameters for each method, Case Study 1.}
	\label{table1}
	\scalebox{1}{\begin{tabular}{|c|c|c|c|c|c|c|c|c|c|}
			\hline
			\multirow{2}{*}{Method} & \multirow{2}{*}{\begin{tabular}[c]{@{}c@{}}Parameter\\ True value\end{tabular}} & \multirow{2}{*}{\begin{tabular}[c]{@{}c@{}}$a_1^*$\\ $0.26$\end{tabular}}      & \multirow{2}{*}{\begin{tabular}[c]{@{}c@{}}$a_2^*$\\ $0.255$\end{tabular}}      & \multirow{2}{*}{\begin{tabular}[c]{@{}c@{}}$a_3^*$\\ $3.125\cdot 10^{-3}$\end{tabular}}     & \multirow{2}{*}{\begin{tabular}[c]{@{}c@{}}$a_4^*$\\ $6.25\cdot 10^{-4}$\end{tabular}}    & \multirow{2}{*}{\begin{tabular}[c]{@{}c@{}}$b_0^*$\\ $4$\end{tabular}}     & \multirow{2}{*}{\begin{tabular}[c]{@{}c@{}}$b_1^*$\\ $0.28$\end{tabular}}   & \multirow{2}{*}{\begin{tabular}[c]{@{}c@{}}$b_2^*$\\ $0.2575$\end{tabular}} & \multirow{2}{*}{\begin{tabular}[c]{@{}c@{}}$b_3^*$\\ $0$\end{tabular}}      \\
			&                                                                                 &                                                                         &                                                                         &                                                                            &                                                &                          &                                                                        &                                                                                                                                                 &                           \\ \hline                                              
			SRIV & MSE  & $3.51\cdot 10^{-4}$  & $8.53\cdot 10^{-5}$  & $1.74\cdot 10^{-8}$ & $5.48\cdot 10^{-10}$ & $2.44\cdot 10^{-2}$ & $6.03\cdot 10^{-3}$ & $1.07\cdot 10^{-4}$ & $1.55\cdot 10^{-7}$ \\ \hline
			SRIVC & MSE  & $3.52\cdot 10^{-4}$  & $7.59\cdot 10^{-5}$  & $1.71\cdot 10^{-8}$ & $4.28\cdot 10^{-10}$ & $2.44\cdot 10^{-2}$ & $4.56\cdot 10^{-3}$ & $1.05\cdot 10^{-4}$ & $0$ \\ \hline
			BCD & MSE & $3.41\cdot 10^{-4}$  & $3.45\cdot 10^{-5}$  & $1.62\cdot 10^{-8}$ & $2.16\cdot 10^{-10}$ & $2.41\cdot 10^{-2}$ & $4.45\cdot 10^{-3}$ & $9.04\cdot 10^{-5}$ & $0$ \\ \hline
	\end{tabular}}
\end{table*}

\subsection{Case Study 2: high-order, highly-resonant system}
\vspace{-0.1cm}
We test the proposed method on a $16$th order system:
\begin{equation}
G(p) = \sum_{i=1}^8 \frac{c_i}{(p/\omega_i)^2 + 2(\xi_i/\omega_i)p + 1}, \notag
\end{equation} 
where the natural frequencies $\omega_i$ are located between $6$ and $470$ [rad/s], and the damping ratios are values between $0.001$ and $0.0017$. The input is a multisine formed by $16$ sinusoids of random phase, and two hundred Monte Carlo runs are recorded with $N=3000$, $h=0.001$[s], and initial conditions equal to zero. Each run includes a zero-mean additive noise of variance equal to 2.25, which corresponds to approximately a signal-to-noise ratio of $20$[dB]. Both SRIVC and BCD methods are initialized at the estimator given by the LSSVF method \citep{young1965determination}. The tolerance factor of each method is set to $\epsilon = 10^{-16}$, and a maximum of $10$ $\mathcal{A}$-iterations with $M_{\textnormal{s}}=200$ is used for BCD.

\begin{figure}
	\begin{minipage}[c]{0.29\linewidth}
		\centering
		\includegraphics[width=\textwidth]{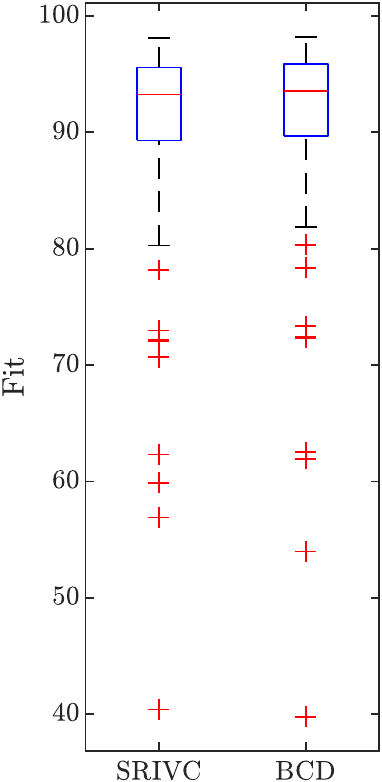}
	\end{minipage}
	\begin{minipage}[c]{0.69\linewidth}
		\centering
		\includegraphics[width=1\textwidth]{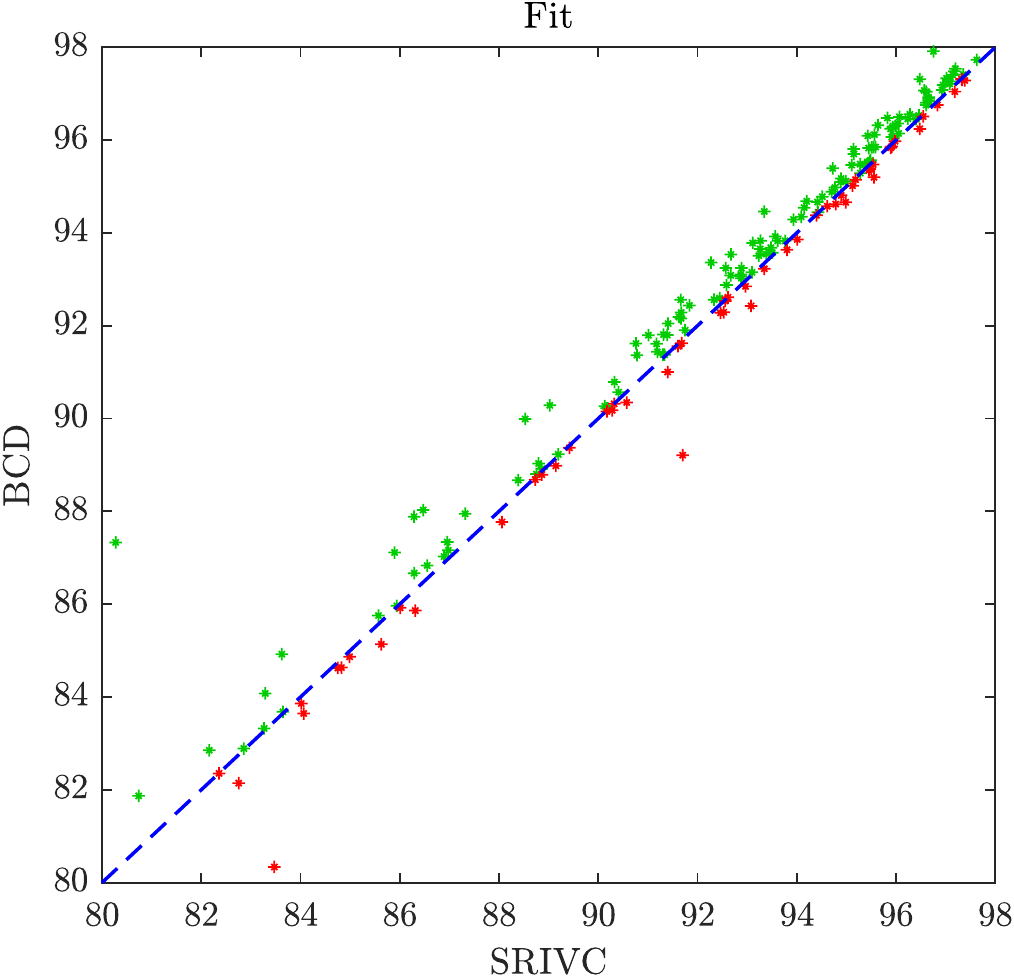}
	\end{minipage}
	\caption{Left figure: Boxplot of the fit metric for the SRIVC and BCD methods, Case Study 2. Right figure: Comparison plot between BCD and SRIVC, Case Study 2. Green dots correspond to runs where BCD outperforms SRIVC. Red dots represent the opposite, and the dashed blue line is the separatrix.}
	\label{fig2}
\end{figure}

Figure \ref{fig2} shows the fit of the SRIVC estimator compared to the proposed BCD method in a boxplot format, and also in a direct comparison plot. In this study, the BCD method returned better fit compared to the SRIVC estimator in $71.5\%$ of the runs, and no run led to a failure in producing a result. In addition, Table \ref{table2} shows the MSEs of the parameters $c_i$ for each approach. The MSEs related to the BCD method are lower or equal to the MSEs of the SRIVC estimates for all the numerator coefficients, which can be explained by the fact that a more parsimonious model is being fit to the data.

\begin{table*}
	\centering
	\caption{MSEs of the estimated numerator parameters $c_i$ for SRIVC and BCD, Case Study 2.}
	\label{table2}
	\scalebox{1}{\begin{tabular}{|c|c|c|c|c|c|c|c|c|c|}
			\hline
			\multirow{2}{*}{Method} & \multirow{2}{*}{\begin{tabular}[c]{@{}c@{}}Parameter\\ True value\end{tabular}} & \multirow{2}{*}{\begin{tabular}[c]{@{}c@{}}$c_1$\\ $0.66$\end{tabular}}      & \multirow{2}{*}{\begin{tabular}[c]{@{}c@{}}$c_2$\\ $0.24$\end{tabular}}      & \multirow{2}{*}{\begin{tabular}[c]{@{}c@{}}$c_3$\\ $0.48$\end{tabular}}     & \multirow{2}{*}{\begin{tabular}[c]{@{}c@{}}$c_4$\\ $0.15$\end{tabular}}    & \multirow{2}{*}{\begin{tabular}[c]{@{}c@{}}$c_5$\\ $0.09$\end{tabular}}     & \multirow{2}{*}{\begin{tabular}[c]{@{}c@{}}$c_6$\\ $0.15$\end{tabular}}   & \multirow{2}{*}{\begin{tabular}[c]{@{}c@{}}$c_7$\\ $0.09$\end{tabular}} & \multirow{2}{*}{\begin{tabular}[c]{@{}c@{}}$c_8$\\ $0.06$\end{tabular}}      \\
			&                                                                                 &                                                                         &                                                                         &                                                                            &                                                &                          &                                                                        &                                                                                                                                                 &                           \\ \hline                                              
			SRIVC & MSE  & $6.52\cdot 10^{-4}$  & $6.74\cdot 10^{-5}$  & $1.61\cdot 10^{-5}$ & $1.13\cdot 10^{-5}$ & $2.19\cdot 10^{-5}$ & $1.04\cdot 10^{-5}$ & $6.06\cdot 10^{-6}$ & $3.87\cdot 10^{-7}$ \\ \hline
			BCD & MSE & $6.33\cdot 10^{-4}$  & $6.42\cdot 10^{-5}$  & $1.61\cdot 10^{-5}$ & $1.09\cdot 10^{-5}$ & $2.14\cdot 10^{-5}$ & $1.04\cdot 10^{-5}$ & $6.00\cdot 10^{-6}$ & $3.84\cdot 10^{-7}$ \\ \hline
	\end{tabular}}
\end{table*}



%
%
\vspace{-0.12cm}
\section{Conclusions}
\label{sec:conclusions}
\vspace{-0.12cm}
In this paper we have derived a method for estimating continuous-time models in an additive form. First, we explored the fact that certain systems might benefit from a more parsimonious model description if a modal form is considered. We proposed a block-coordinate descent method to identify models with this structure, and we proved its convergence. The numerical simulations show that the method delivers more accurate models than the standard indirect and direct methods, and its potential for the estimation of large-order and highly-resonant systems has been studied.

\vspace{-0.12cm}
\bibliography{References}
\vspace{-0.12cm}
\appendix 
\section{Proof of Theorem \ref{thm42}}
\vspace{-0.13cm}
\label{appendixthm42}
\begin{pf}
We adapt the theory in Section 7.7 of \cite{luenberger2008linear} to first analyze the global convergence of Algorithm \ref{algorithm1} to the solution set $\tilde{\Gamma}_N = \{ \mathbf{x} \in \Omega \colon \mathcal{A}(\mathbf{x}) = \mathbf{x}\}$. Afterwards, we prove that $\tilde{\Gamma}_N\subseteq \Gamma_N$. By Lemma B.1 of \cite{soderstrom1989system} (see also \cite{soderstrom1975ergodicity}) $V_N(\mathbf{x})$ converges almost surely (as $N\to\infty$) to
\begin{align}
	\label{barv}
	\bar{V}\hspace{-0.03cm}(\mathbf{x})\hspace{-0.1cm} &= \hspace{-0.02cm}\mathbb{E}\hspace{-0.02cm}\left\{\hspace{-0.03cm}\left[ y(kh) \hspace{-0.05cm}-\hspace{-0.05cm} \sum_{i=1}^K G_i(p,\bm{\theta}_i) u(kh) \right]^{\hspace{-0.03cm}2}\right\} \\
	&= \hspace{-0.08cm}\frac{1}{2 \pi} \hspace{-0.09cm}\int_{\hspace{-0.03cm}-\hspace{-0.03cm}\frac{\pi}{h}}^{\frac{\pi}{h}} \hspace{-0.06cm}\left|\sum_{i=1}^K \hspace{-0.07cm}\left[\hspace{-0.02cm}\tilde{G}_{\hspace{-0.02cm}i}^*\hspace{-0.03cm}(\hspace{-0.02cm}e^{i \omega h}\hspace{-0.02cm})\hspace{-0.07cm}-\hspace{-0.05cm}\tilde{G}_{\hspace{-0.02cm}i}(\hspace{-0.02cm}e^{i \omega h}\hspace{-0.03cm},\hspace{-0.02cm}\bm{\theta}_i)\right] \hspace{-0.03cm}\right|^2 \hspace{-0.14cm}\Phi_u(\omega) \textnormal{d}\omega \hspace{-0.05cm}+\hspace{-0.05cm} \sigma^2\hspace{-0.04cm}, \notag
\end{align}
where $\Phi_u$ is the spectrum of the sampled input signal, and $\tilde{G}_i^*$, $\tilde{G}_i$ are the discrete-time ZOH equivalents of the system $G_i^*$ and model $G_i$, respectively. Since $V_N$ is continuous and $\Omega$ is compact, this convergence is uniform. In the following, we prove the necessary ingredients for applying the Global Convergence Theorem of \cite{luenberger2008linear}.

\begin{itemize}
	\item \emph{Well-posedness of $\mathbf{S}$}: Due to the uniform convergence of $V_N$ to $\bar{V}$, for almost every realization $\xi$, there exists a sample size $N'(\xi) \in \mathbb{N}$ such that for all $\mathbf{x} = [\bm{\theta}_1^\top, \; \dots, \; \bm{\theta}_K^\top]^\top \in \Omega$ and $N \geq N'(\xi)$, the minimizer of $V(\bm{\theta}_1, \dots, \bm{\theta}_{i-1}, \bm{\theta}$, $\bm{\theta}_{i+1}, \dots, \bm{\theta}_K)$ with respect to $\bm{\theta}$ is unique. Thus, for all $N \geq N'(\xi)$, $\mathbf{S}$ is a well-defined point-to-point mapping.
	
	\item \emph{Descent of algorithm (with respect to $V_N$)}: From the definition of the algorithm $\mathcal{A}$ in \eqref{algorithma} and the solution set $\tilde{\Gamma}_N$, it follows that if $\mathbf{x} \in \tilde{\Gamma}_N$, then $V_N(\mathcal{A}(\mathbf{x})) \leq V_N(\mathbf{x})$. Otherwise, if $\mathbf{x} \notin \tilde{\Gamma}_N$, then $\mathbf{x}\neq \mathcal{A}(\mathbf{x})$ and therefore we must have $V_N(\mathcal{A}(\mathbf{x})) < V_N(\mathbf{x})$. Thus, $V_N$ is a continuous \emph{descent function} for $\tilde{\Gamma}_N$ and $\mathcal{A}$.
	
	\item \emph{Closedness of $\mathcal{A}$}: From the theorem in Section 8.4 of \cite{luenberger2008linear}, the map $\mathbf{S}$ is closed. In addition, the maps $\mathbf{C}^i$ ($i = 1, \dots, n$) are continuous, and thus closed. Therefore, by Corollary 1 in Section 7.7 of \cite{luenberger2008linear}, $\mathcal{A}$ is closed.
\end{itemize}

From the previous points, it follows that all the conditions for the Global Convergence Theorem in Section 7.7 of \cite{luenberger2008linear} hold. This implies that, for almost every realization and for sufficiently large $N$, the limit of every converging subsequence of $(\mathbf{x}^l)_{l \in \mathbb{N}}$ belongs to the solution set $\tilde{\Gamma}_N$.

Now, to prove that $\tilde{\Gamma}_N\subseteq \Gamma_N$, take $\bar{\mathbf{x}}= [\bar{\bm{\theta}}_1^\top, \dots, \bar{\bm{\theta}}_K^\top]^\top\in\tilde{\Gamma}_N$. Since $\bar{\bm{\theta}}_1$ is an critical point of the function $V_N^1\hspace{-0.02cm}(\bm{\theta}_1):=V_N(\bm{\theta}_1,\bar{\bm{\theta}}_2, \dots, \bar{\bm{\theta}}_K)$ for fixed $\bar{\bm{\theta}}_2, \dots, \bar{\bm{\theta}}_K$, then $V_N^1$ must satisfy the first-order optimality condition $\frac{\partial V_N^1}{\partial \bm{\theta}_1}|_{\bm{\theta}_1=\bar{\bm{\theta}}_1} = \mathbf{0}$. By definition, this means that $\frac{\partial V_N}{\partial \bm{\theta}_1}|_{\bm{\theta}_1=\bar{\bm{\theta}}_1} = \mathbf{0}$. Repeating this argument for $\bar{\bm{\theta}}_2, \dots,\bar{\bm{\theta}}_K$ leads to the desired conclusion, namely, that $\bar{\mathbf{x}}\in \Gamma_N$. \hspace{4.9cm} \hfill \qed 
\end{pf}

\vspace{-0.1cm}
\section{Proof of Lemma \ref{lemmagnsrivc}}
\label{prooflemmagn}	
\vspace{-0.13cm}
\begin{pf}
As $s\to \infty$, any converging point $\bm{\theta}_i^{l+1}:=\lim_{s\to \infty} \bm{\theta}_{i,s}^{l+1}$ of the SRIVC iterations in \eqref{formulasrivc} must satisfy
\begin{align}
	\bm{\theta}_{i}^{l+1} &= \left[\frac{1}{N}\sum_{k=1}^{N}\hat{\bm{\varphi}}_\textnormal{f}(kh,\bm{\theta}_{i}^{l+1}) \bm{\varphi}_\textnormal{f}^\top(kh,\bm{\theta}_{i}^{l+1})  \right]^{-1} \notag \\
	&\hspace{0.4cm}\times \hspace{-0.1cm} \left[\frac{1}{N}\hspace{-0.05cm}\sum_{k=1}^{N}\hspace{-0.06cm}\hat{\bm{\varphi}}_\textnormal{f}(kh,\hspace{-0.02cm}\bm{\theta}_{i}^{l\hspace{-0.02cm}+\hspace{-0.02cm}1}) \tilde{y}_{\textnormal{f}}(kh,\bm{\theta}_{i}^{l+1}) \right], \notag
\end{align}
which, given the non-singularity of the normal matrix above, is equivalent to
\begin{equation}
	\label{conditionsrivc}
	\hspace{-0.3cm}\frac{1}{N}\hspace{-0.07cm}\sum_{k=1}^N \hspace{-0.07cm}\hat{\bm{\varphi}}_\textnormal{f}(\hspace{-0.02cm}kh,\hspace{-0.02cm}\bm{\theta}_i^{l\hspace{-0.02cm}+\hspace{-0.02cm}1}\hspace{-0.03cm}) \hspace{-0.07cm}\left[\tilde{y}_\textnormal{f}(\hspace{-0.02cm}kh,\hspace{-0.02cm}\bm{\theta}_i^{l\hspace{-0.02cm}+\hspace{-0.02cm}1}\hspace{-0.03cm})\hspace{-0.08cm}-\hspace{-0.06cm}\bm{\varphi}_\textnormal{f}^{\hspace{-0.05cm}\top}\hspace{-0.08cm}(\hspace{-0.01cm}kh,\hspace{-0.02cm}\bm{\theta}_i^{l\hspace{-0.02cm}+\hspace{-0.02cm}1}\hspace{-0.03cm})\bm{\theta}_i^{l\hspace{-0.02cm}+\hspace{-0.02cm}1}\hspace{-0.02cm} \right]\hspace{-0.13cm} =\hspace{-0.07cm} \mathbf{0}.\hspace{-0.137cm}
\end{equation}
However, by leveraging the expressions for $\bm{\varphi}_\textnormal{f}$ and $\tilde{y}_\textnormal{f}$ in \eqref{filteredregressor} and \eqref{filteredoutput}, we find that
\begin{align}
	\tilde{y}_\textnormal{f}&(\hspace{-0.01cm}kh,\hspace{-0.02cm}\bm{\theta}_i^{l\hspace{-0.02cm}+\hspace{-0.02cm}1}\hspace{-0.02cm})\hspace{-0.08cm}-\hspace{-0.06cm}\bm{\varphi}_\textnormal{f}^{\hspace{-0.04cm}\top}\hspace{-0.07cm}(\hspace{-0.01cm}kh,\hspace{-0.02cm}\bm{\theta}_i^{l\hspace{-0.02cm}+\hspace{-0.02cm}1}\hspace{-0.02cm})\bm{\theta}_i^{l\hspace{-0.02cm}+\hspace{-0.02cm}1}\notag \\
	&=\frac{1}{A_i^{l\hspace{-0.02cm}+\hspace{-0.02cm}1}(p)}\tilde{y}(\hspace{-0.01cm}kh\hspace{-0.01cm})\hspace{-0.07cm}-\hspace{-0.08cm}\frac{(1\hspace{-0.08cm}-\hspace{-0.09cm}A_i^{l\hspace{-0.02cm}+\hspace{-0.02cm}1}(p)\hspace{-0.02cm})}{A_i^{l\hspace{-0.02cm}+\hspace{-0.02cm}1}(p)}\tilde{y}(\hspace{-0.01cm}kh\hspace{-0.01cm})\hspace{-0.09cm}-\hspace{-0.09cm}\frac{B_i^{l\hspace{-0.02cm}+\hspace{-0.02cm}1}(p)}{A_i^{l\hspace{-0.02cm}+\hspace{-0.02cm}1}(p)}u(\hspace{-0.01cm}kh\hspace{-0.01cm}) \notag \\
	&=e(kh,\bm{\theta}_i^{l\hspace{-0.02cm}+\hspace{-0.02cm}1}). \notag
\end{align}
Thus, replacing this result into \eqref{conditionsrivc} also leads to \eqref{convergingto}, concluding the proof. \hspace{4.9cm}\hfill \qed 
\end{pf}

\end{document}